\documentclass[conference]{IEEEtran}
\IEEEoverridecommandlockouts

\usepackage{algorithmic}
\usepackage{graphicx}
\usepackage{textcomp}
\usepackage{xcolor}
\usepackage{amsmath, calc}
\usepackage{tcolorbox}
\usepackage{multirow}
\usepackage{url}
\usepackage{hyperref}
\usepackage{pgfplots}
\pgfplotsset{compat=1.18}
\usepackage{placeins}
\usepackage{fontawesome}
\usepackage{comment}
\usepackage{tabularx}
\usepackage{enumitem}
\usepackage{booktabs}
\usepackage{framed}
\usepackage{fancybox}
\usepackage{makecell}
\usepackage{csquotes}
\usepackage{orcidlink}
\usepackage{multicol}
\usepackage{colortbl}
\usepackage{caption}

\def\BibTeX{{\rm B\kern-.05em{\sc i\kern-.025em b}\kern-.08em
    T\kern-.1667em\lower.7ex\hbox{E}\kern-.125emX}}
    
\begin{document}

\title{Towards Trustworthy Sentiment Analysis in Software Engineering: Dataset Characteristics and Tool Selection}

\makeatletter
\newcommand{\linebreakand}{%
  \end{@IEEEauthorhalign}
  \hfill\mbox{}\par
  \mbox{}\hfill\begin{@IEEEauthorhalign}
}
\makeatother


\author{
    \IEEEauthorblockN{
        Martin Obaidi\orcidlink{0000-0001-9217-3934},
        Marc Herrmann\orcidlink{0000-0002-3951-3300},
        Kurt Schneider\orcidlink{0000-0002-7456-8323}
    }
    \IEEEauthorblockA{
        \textit{Leibniz University Hannover, Software Engineering Group} \\
        Hannover, Germany \\
        \{martin.obaidi, marc.herrmann, kurt.schneider\}@inf.uni-hannover.de
    }
    \and
    \IEEEauthorblockN{
        Jil Klünder\orcidlink{0000-0001-7674-2930}
    }
    \IEEEauthorblockA{
        \textit{University of Applied Sciences} \\
        \textit{FHDW Hannover} \\
        Hannover, Germany \\
        jil.kluender@fhdw.de
    }
}

\maketitle

\begin{abstract}
Software development relies heavily on text-based communication, making sentiment analysis a valuable tool for understanding team dynamics and supporting trustworthy AI-driven analytics in requirements engineering. However, existing sentiment analysis tools often perform inconsistently across datasets from different platforms, due to variations in communication style and content.

In this study, we analyze linguistic and statistical features of 10 developer communication datasets from five platforms and evaluate the performance of 14 sentiment analysis tools. Based on these results, we propose a mapping approach and questionnaire that recommends suitable sentiment analysis tools for new datasets, using their characteristic features as input.

Our results show that dataset characteristics can be leveraged to improve tool selection, as platforms differ substantially in both linguistic and statistical properties. While transformer-based models such as SetFit and RoBERTa consistently achieve strong results, tool effectiveness remains context-dependent. Our approach supports researchers and practitioners in selecting trustworthy tools for sentiment analysis in software engineering, while highlighting the need for ongoing evaluation as communication contexts evolve.
\end{abstract}



\section{Introduction}
\label{sec:intro}

Sentiment analysis is an increasingly important tool in software engineering (SE) \cite{obaidi2021development,obaidiSentiSMS22,linSentiSLR22}, as it supports understanding and improving team communication and collaboration \cite{obaidi2022smszenodo}. Modern SE relies on textual communication, e.g., in issue trackers or chat tools such as Slack \cite{Novielli.2018b,Calefato.2018}, prompting the development of sentiment analysis tools tailored to developer interactions \cite{Calefato.2018,herrmann-kluender-2021,Islam.2018,Ahmed.2017, herrmann2021automatic, obaidi2025GoldStandardDE}. These tools are typically based on machine learning and are trained using datasets labeled with sentiment polarity (positive, negative, neutral). Such datasets reflect technical language and platform-specific conventions, which differ significantly from everyday user-generated text \cite{specht2024sentiment,madampe20agilesentiment}. Understanding the nuances of developer communication has become integral to software process analysis, especially as teams adopt new collaborative practices and AI-powered tooling.

A major challenge is the variability of tool performance across datasets and contexts \cite{noviellicross20,obaidi22cross,9240704,10.1145/3180155.3180195}. Tools trained on one platform (e.g., GitHub) often perform poorly when applied to data from others (e.g., Jira), and combining datasets or tools has not consistently yielded better results \cite{obaidi22cross,uddinVoting2021}. Lexicon-based approaches can be more robust but lack domain nuance \cite{noviellicross20}. These performance differences are closely linked to dataset characteristics, such as specific linguistic features (e.g., politeness, technical focus) and statistical attributes (e.g., document length, emoticon frequency) \cite{obaidi22cross,CABRERADIEGO2020105633}. For instance, Stack Overflow posts are often concise and factual, while Jira discussions tend to be more detailed and collaborative \cite{CABRERADIEGO2020105633}.

For researchers and practitioners seeking to analyze new, unlabeled communication data (e.g., \cite{10.1145/3392877,Mostafa.2018,madampe20agilesentiment}), the question arises: Which sentiment analysis tool is most appropriate for the dataset at hand? Comparing dataset characteristics to those of existing, labeled datasets can help to inform this choice, especially as retraining or building new tools is often infeasible.

Despite the relevance of this challenge, no prior work has systematically mapped the linguistic and statistical characteristics of SE datasets to the performance of sentiment analysis tools, or provided practical guidance for tool selection in the face of dataset variability. In this paper, we address this gap with the following contributions:
\begin{itemize}
    \item An analysis of ten labeled datasets from five SE platforms, identifying key linguistic and statistical characteristics relevant for sentiment analysis.
    \item A comprehensive evaluation of 14 sentiment analysis tools, reporting their performance on each dataset.
    \item A mapping between dataset characteristics and tool performance, enabling informed recommendations for tool selection.
    \item A questionnaire that supports practical tool recommendations for new, unlabeled datasets based on their communication characteristics.
\end{itemize}

Our approach enables practitioners and researchers to select sentiment analysis tools that are well-matched to their data, supporting more trustworthy and explainable AI-based analysis of software communication. Rather than requiring detailed linguistic annotation of new datasets, our questionnaire allows users to make informed, subjective assessments based on empirically derived features.

While our current mapping uses platforms as reference points, the underlying method is based on dataset characteristics, meaning the questionnaire output can be seen as a ``profile'' that describes a communication style rather than prescribing a strict platform. Future extensions could make more use of this characteristic-based matching, e.g., via clustering or profiling techniques, to further decouple recommendations from platform names.

The remainder of this paper is structured as follows: Section~\ref{chap:related} reviews related work. Section~\ref{chap:study} presents our study design. Section~\ref{chap:results} reports our results, and Section~\ref{chap:discussion} discusses implications and limitations. Section~\ref{chap:conclusion} concludes.

\section{Background and Related Work}
\label{chap:related}

Sentiment analysis has become central in software engineering (SE) research, enabling the study of communication, collaboration, and well-being in development teams~\cite{linSentiSLR22,obaidi2021development,obaidiSentiSMS22}. In this section, we discuss prior work on (a) sentiment datasets and their characteristics, (b) sentiment analysis tools in SE, and (c) recent findings regarding subjectivity and trustworthy interpretation in team settings.

\subsection{Datasets for Sentiment Analysis in SE}
A range of labeled datasets for sentiment analysis in SE have been developed, each reflecting the linguistic and contextual nuances of different platforms. For instance, Stack Overflow datasets~\cite{Novielli.2018b,8643972,10.1145/3180155.3180195} typically consist of concise, technical posts labeled for sentiment polarity by multiple raters, while datasets from GitHub~\cite{novielligold.2020,10.1145/3194932.3194935,Imtiaz.2018} capture a mix of issue comments and code reviews. Jira datasets~\cite{Ortu.2016,Islam.2018b} focus on collaborative problem solving and are often labeled with both emotions and polarity. Code review datasets~\cite{Ahmed.2017,Imtiaz.2018} and app review datasets~\cite{10.1145/3180155.3180195} reflect further diversity in communication style, granularity, and annotation schemes.

Dataset construction practices vary significantly: some use emotion annotation later mapped to sentiment polarity, while others rely directly on polarity labels. Annotation is often performed by multiple raters, but inter-rater agreement remains a challenge~\cite{herrmannSentiSurvey22}. Moreover, platform-specific jargon and evolving communication styles complicate generalization across datasets. The application of sentiment analysis in industrial settings has recently attracted attention, highlighting the practical challenges and potential for real-world use~\cite{specht2024sentiment,schroth2022potential}.

\subsection{Sentiment Analysis Tools in SE}
Early sentiment analysis in SE relied on lexicon-based or classical machine learning tools, such as SentiStrength~\cite{Thelwall.2012}, SentiStrength-SE~\cite{Islam.2018}, and Senti4SD~\cite{Calefato.2018}. These tools incorporate domain-specific dictionaries or heuristics, enabling polarity classification of developer communication without the need for retraining. More recent approaches employ supervised machine learning and deep learning: SentiCR~\cite{Ahmed.2017} and SentiSW~\cite{10.1145/3194932.3194935} use features such as TF-IDF and gradient boosting, while transformer-based models like BERT and RoBERTa have demonstrated strong results when fine-tuned on SE data~\cite{biswas2020sentimentbert,uddinVoting2021}. Novel techniques also leverage emoji-based supervision~\cite{Chen2019sentimoji} or ensemble/voting schemes~\cite{herrmann-kluender-2021,sentianalyzerreport2022}.

Recent benchmarks highlight the continued evolution in this field, with large language models (LLMs) such as GPT and zero-shot/few-shot learning approaches increasingly evaluated for sentiment analysis in SE~\cite{Zhang2024Revisiting}. However, tool performance remains highly sensitive to dataset characteristics and annotation schemes, motivating the need for more principled recommendations.

\subsection{Trustworthiness, Subjectivity, and Sentiment Perception}
A recurring challenge is the subjectivity of sentiment annotation and perception among developers. Recent studies reveal substantial variability in human sentiment judgments: Herrmann et al.~\cite{herrmann2025different-perceptions} found distinct developer groups with differing perceptions, independent of demographics, and showed via simulation~\cite{herrmann2025montecarlo} that incomplete team mood surveys can distort aggregate sentiment. Annotation agreement with gold labels in SE sentiment datasets can be as low as 62.5\%~\cite{herrmannSentiSurvey22,obaidi2022sentisurveyzenodo}, underlining the difficulty of obtaining reliable ground truth. In addition, the need for explainability and adaptation to subjective user requirements has been discussed in the context of SE~\cite{obaidi2025mood}.

These findings emphasize the need for trustworthy, explainable, and context-aware sentiment analysis in SE,an essential goal when integrating such tools in critical, human-centric AI pipelines. Tool evaluation and recommendations must account for both the technical characteristics of datasets and the inherent variability in human perception.

\subsection{Gaps and Motivation}
While many tools and datasets exist, few studies systematically analyze how linguistic and statistical dataset characteristics affect tool performance, or offer practical guidance for tool selection across diverse contexts. In particular, the mapping between dataset properties and trustworthy tool recommendations remains an open research challenge, with significant implications for explainable AI and requirements engineering in SE.

\section{Study Design}
\label{chap:study}

This section outlines our selection of tools and datasets, the approach for analyzing dataset characteristics and tool performance, and our mapping procedure for tool recommendation. Figure~\ref{fig:approach} provides an overview.

\begin{figure}[htb!]
\includegraphics[width=\columnwidth]{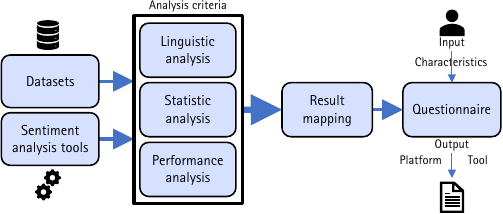}
\caption{Overview of the study design.}
\label{fig:approach}
\end{figure}

To recommend suitable sentiment analysis tools for new software engineering (SE) datasets, we first analyze linguistic and statistical characteristics of diverse datasets, then assess the performance of sentiment analysis tools, and finally map these insights to a practical questionnaire for tool recommendation.

\subsection{Research Questions}
\label{chap:researchquestins}

Our work is guided by the following research questions:
\\
\textbf{RQ1:} \textit{How do SE communication datasets differ in their linguistic and statistical characteristics?}
\\
\textbf{RQ2:} \textit{How do sentiment analysis tools perform on communication data from different SE platforms?}
\\
\textbf{RQ3:} \textit{What is the best sentiment analysis tool for a given, unlabeled communication dataset based on its characteristics?}

\subsection{Data}
\label{chap:data}
We selected sentiment analysis datasets from a recent systematic literature review~\cite{linSentiSLR22}, focusing on datasets with polarity labels (positive, negative, neutral). If emotion labels could be reliably mapped to polarity using established heuristics~\cite{noviellicross20}, we included them as well. Table~\ref{tab:datasets} summarizes the datasets and their class distributions.

\begin{table}[htbp!]
\footnotesize
\caption{Overview of the used datasets. The symbol "\#" means "number of" and "Docs" stands for "documents". *The class neutral in Code Reviews is "non-negative". n stands for the sample size with equal or greater than 95\% confidence interval and an error rate of 5\%.}
\label{tab:datasets}
\begin{tabularx}{\columnwidth}{@{}Xlrrrrr@{}}
\toprule
\textbf{Platform}         & \textbf{Dataset}    & \textbf{\#Docs} & \textbf{\#Neg} & \textbf{\#Neu} & \textbf{\#Pos} & \textbf{$n$}\\ \midrule
App Reviews                   & App~\cite{10.1145/3180155.3180195}      & 341    & 130   & 25    & 186  & 193 \\\midrule
Code Reviews                  & Code~\cite{Ahmed.2017}                  & 1600   & 398   & 1202* & 0* & 310   \\\midrule
\multirow{3}{*}{GitHub}         & GH 1~\cite{novielligold.2020}               & 7122   & 2087  & 3022  & 2013 & - \\
                                     & GH 2~\cite{10.1145/3194932.3194935}         & 2962   & 401   & 1968  & 593 & -   \\
                                     & GH 3~\cite{Imtiaz.2018}                    & 585    & 73    & 419   & 93 & -    \\
                                     & $\sum$ & 10669 & 2561 & 5409 & 2699 & 371\\\midrule
\multirow{2}{*}{Jira}                & Jira 1~\cite{Islam.2018}                        & 1795   & 541   & 616   & 638   \\
                                     & Jira 2~\cite{Ortu.2016}                        & 6300   & 778   & 4165  & 1357 & -   \\
                                     & $\sum$ & 8095 & 1319 & 4781 & 1995 & 367\\\midrule
\multirow{3}{*}{Stack Overflow} & SO 1~\cite{10.1145/3180155.3180195} & 1500   & 178   & 1191  & 131  & -  \\
                                     & SO 2~\cite{Novielli.2018b}          & 4687   & 1145  & 1947  & 1595 & -  \\
                                     & SO 3~\cite{8643972}                 & 4518   & 496   & 3132  & 890 & -  \\
                                     & $\sum$ & 10705 & 1819 & 6270 & 2616 & 371\\
                                     \bottomrule
\end{tabularx}
\end{table}

When necessary, we mapped emotion annotations to polarity using heuristics from prior work~\cite{noviellicross20}, while acknowledging potential limitations. For instance, in Jira datasets, emotions such as "Excited" were mapped to positive, while "Stress" was mapped to negative. In ambiguous cases, only documents with clear mapping were included. The neutral class was operationalized as documents with neutral or no strong valence.

For each platform, all datasets were pooled, and a representative random sample was drawn (minimum sample size for 95\% confidence and 5\% error). Sampling preserved original polarity distributions except for App Reviews, where all neutral items were retained for adequate class representation.

\subsubsection{Data Pre-processing}
Datasets were pre-processed to remove entries irrelevant for polarity classification (e.g., sarcasm in GH 3~\cite{Imtiaz.2018}), to harmonize column structures, and to map emotion labels where needed. Details of initial processing can be found in the respective dataset papers.

\subsection{Sentiment Analysis Tools}
\label{chap:tools}
We selected 14 representative sentiment analysis tools, covering transformer models (e.g., BERT~\cite{Devlin.2019bert}, RoBERTa, ELECTRA, SetFit~\cite{tunstall2022setfit}, XLNet) and established SE-specific tools (e.g., Senti4SD~\cite{Calefato.2018}, SentiCR~\cite{Ahmed.2017}, SentiStrength~\cite{Thelwall.2012}, SentiStrength-SE~\cite{Islam.2018}, SentiSW~\cite{10.1145/3194932.3194935}, SEnti-Analyzer~\cite{sentianalyzerreport2022}). Both lexicon-based and machine learning methods are included for breadth.

Tools were used as described in their respective publications, with adaptation to our class labels if needed. For tools with pre-defined train/test splits (e.g., SentiCR, Senti4SD), we avoided data leakage by retraining solely on our training splits (80/20), never reusing the original splits.

\subsection{Performance Metrics}
\label{chap:metrics}
We evaluated tool performance using micro- and macro-averaged $F_1$ scores, as well as their average (overall score), following established practice~\cite{noviellicross20,9240704}. The micro $F_1$ reflects accuracy, while macro $F_1$ balances across classes. This dual perspective mitigates the impact of class imbalance. Performance is reported under consistent, controlled conditions, and findings are interpreted as empirical guidance rather than definitive rankings, given the inherent context-sensitivity of sentiment classification in SE. While $F_1$ is widely used, we note that in certain applications, other metrics such as precision, recall, or area under the curve may be preferable depending on project goals. Thus, our evaluation provides a baseline, but future work could extend this to support alternate validation objectives.

\subsection{Platform Analysis Approach}
\label{chap:approach}

\subsubsection{Inputs}

Our analysis is based on the datasets (Section~\ref{chap:data}) and sentiment analysis tools (Section~\ref{chap:tools}). Each dataset was analyzed for linguistic features and statistical values, and each tool was evaluated on all datasets for performance.

\subsubsection{Linguistic Features}

We employed a multi-step, collaborative coding process to identify linguistic features characteristic of SE communications across platforms (see Figure~\ref{fig:codingB}). Initially, three researchers jointly coded a small, balanced sample (15 positive, 15 neutral, 15 negative statements per platform), iteratively refining a shared feature list. Then, each researcher independently coded additional samples (10 per polarity per platform), updating the feature set as needed. 

All identified linguistic features were subsequently grouped into categories. The full platform samples were then labeled by all three researchers, enabling calculation of Fleiss' $\kappa$ for interrater reliability and resolving disagreements via majority vote. For each platform, the frequency of each linguistic feature was computed as the percentage of documents exhibiting that feature.

\begin{figure}[htb!]
\includegraphics[width=0.975\columnwidth]{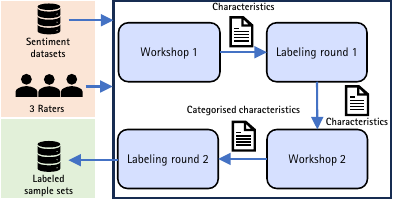}
\caption{Steps of linguistic feature coding and platform analysis.}
\label{fig:codingB}
\end{figure}

\subsubsection{Statistics}

In addition to linguistic features, we computed eight statistical values for each sample: (1) average number of characters per document, (2) characters per word, (3) words per document, (4) capitalized words, (5) spelling mistakes (via \texttt{pyspellchecker}), (6) emoticons, (7) question marks, and (8) exclamation marks. These metrics are standard in prior work~\cite{10.1007/978-3-030-64266-2_8,Claes.2018} and provide further insight into dataset characteristics.

\subsubsection{Performance}

For tool evaluation, each dataset was split into 80\% training and 20\% test sets (for ML-based tools). Dictionary-based tools were evaluated on the full test set. We computed micro- and macro-averaged $F_1$ scores, and their mean (overall score), using the \texttt{classification\_report} from Scikit-learn\footnote{\url{https://scikit-learn.org/stable/modules/generated/sklearn.metrics.classification_report.html}}. This process allowed us to empirically identify the best-performing tool for each dataset, supporting our final tool recommendation mapping.

\subsubsection{Result Mapping \& Questionnaire}
\label{chap:questionnaire}

After analyzing dataset characteristics and tool performance, we map these results to user input via a structured questionnaire. This mapping is essential, as communication styles evolve and may differ even within the same platform. While our mapping currently uses platforms as references for characteristic profiles, the underlying approach analyzes datasets based on their linguistic and statistical features. Thus, the resulting mapping can also be interpreted as matching to a characteristic profile, enabling future extensions that cluster or match datasets beyond strict platform labels.

Our approach enables users to compare their own (potentially new or proprietary) datasets to existing SE communication data, supporting robust tool recommendations even in cases of dataset drift or domain adaptation. This also enhances the trustworthiness and transparency of the tool selection process by grounding recommendations in explainable, empirically derived characteristics.

\paragraph{Mapping Linguistic Features.}
Linguistic feature frequencies (as percentages) are divided into four intervals, each corresponding to a Likert-style answer (see Table~\ref{tab:answers}). Users indicate, for each linguistic feature, whether it is “Fully true,” “More likely to be true,” “More unlikely to be true,” or “Not true at all” for their dataset. This supports subjective, yet structured, mapping without requiring users to calculate exact frequencies.

\begin{table}[t]
\footnotesize
\caption{Answer options and corresponding frequency intervals}
\label{tab:answers}
\begin{tabularx}{\columnwidth}{@{}XX@{}}
\toprule
\textbf{Answer Option}     & \textbf{Interval}   \\ \midrule
Fully true (True)                   & 75--100\%           \\
More likely to be true (Likely)     & 50--75\%            \\
More unlikely to be true (Unlikely) & 25--50\%            \\
Not true at all (Untrue)            & 0--25\%             \\
Not specified                       & ---                 \\ \bottomrule
\end{tabularx}
\end{table}

Each platform is mapped to an interval per feature. If a user’s answer matches the interval for a platform, that platform receives a point. If no platform fits the interval or the user selects “Not specified,” the feature is mapped as “Ambiguous.”

\paragraph{Mapping Statistical Values.}
Since statistical values (e.g., average words per document) are absolute numbers, we calculate the absolute difference between the user’s input and each platform’s value, assigning a point to the platform with the smallest difference for each statistic. This ensures a flexible matching even for out-of-distribution cases.

\paragraph{Recommendation Output.}
After mapping all features and statistics, the platform with the highest score is recommended, along with its empirically best-performing sentiment analysis tool. In the case of a tie, all tied platforms and corresponding tools are recommended. If most answers are “Not specified” or if no platform matches the majority of features, the dataset is considered “Ambiguous.” In this case, we recommend both SetFit~\cite{tunstall2022setfit} and the dictionary-based SentiStrength-SE~\cite{Islam.2018}, which demonstrated robust, cross-platform performance.

This approach allows practical, low-barrier tool selection without requiring in-depth NLP expertise, and supports robust handling of uncertainty or novel datasets—aligning with trustworthy, adaptable AI practices.

\subsection{Interrater Agreement}
\label{sub:interrater}

To assess the reliability of our manual linguistic feature coding, we computed Fleiss' $\kappa$~\cite{fleiss1971measuring} to quantify agreement among the three raters (researchers). Agreement was interpreted using the categories from Landis and Koch~\cite{landis1977measurement}. For additional transparency, we report both $\kappa$ and raw agreement percentages (e.g., 60\% agreement if raters agree on 6 out of 10 items).

\section{Results}
\label{chap:results}

This section presents the results of our platform and tool performance analyses.

\subsection{Linguistic Feature Analysis}
\label{chap:characteristicanalysis}

We identified 13 linguistic features relevant to SE communication, summarized in Table~\ref{tab:characteristics}. Each document in the sample sets was coded for the presence or absence of these features.

\begin{table}[htbp!]
\footnotesize
\setlength{\tabcolsep}{2pt}
\caption{Linguistic features identified in SE communication}
\label{tab:characteristics}
\begin{tabularx}{\columnwidth}{lX}
\toprule
\textbf{Feature} & \textbf{Description} \\ \midrule
1. Direct emotion & Explicit expression of emotions \\
2. Emphasized positivity & Strongly expressed positive opinions \\
3. Technical focus & Predominantly technical content \\
4. Balanced critique & Negative comments with positive points included \\
5. Progress sharing & Reporting on project progress or achievements \\
6. Gratitude & Expressing thanks (distinct from compliments) \\
7. Inquisitive & Asking about code-level decisions \\
8. Help seeking/offering & Requests for, or offers of, help \\
9. Compliments & Compliments on others' work (distinct from gratitude) \\
10. Bug fix requests & Explicit requests for bug fixes \\
11. Constructive criticism & Negative feedback with justification or advice \\
12. Username mentioning & Mentioning users by handle \\
13. Name mentioning & Mentioning users by name \\ \bottomrule
\end{tabularx}
\end{table}

Features 1, 2, 6, and 9 relate to emotion and opinion expression; 4 and 11 to feedback dynamics; 3, 5, and 7 to content nature; 8 and 10 to assistance and problem-solving; and 12/13 to personal addressing.

\paragraph{Interrater Agreement.}
Table~\ref{tab:fleiss} summarizes interrater reliability for coding the 13 features, as measured by Fleiss’ $\kappa$ and raw agreement. Across all platforms and features, agreement ranged from substantial to almost perfect, supporting the reliability of the coding process.

\begin{table}[htbp!]
\footnotesize
\setlength{\tabcolsep}{4pt}
\caption{Interrater agreement: Fleiss' $\kappa$ and agreement rates for feature coding}
\label{tab:fleiss}
\begin{tabular}{lrrrrrrrrrr}
\toprule
\textbf{Ling}    & \multicolumn{2}{c}{App} & \multicolumn{2}{c}{Code} & \multicolumn{2}{c}{GH} & \multicolumn{2}{c}{Jira} & \multicolumn{2}{c}{SO} \\  \midrule
& $\kappa$ & Agr & $\kappa$ & Agr & $\kappa$ & Agr & $\kappa$ & Agr & $\kappa$ & Agr \\ \midrule
$L_1$  & 0.84 & 0.89 & 0.90 & 0.98 & 0.89 & 0.96 & 0.88 & 0.96 & 0.88 & 0.97 \\
$L_2$  & 0.81 & 0.88 & 1.00 & 1.00 & 0.79 & 0.91 & 0.84 & 0.98 & 0.89 & 0.98 \\
$L_3$  & ---  & 1.00 & 0.82 & 0.86 & 0.76 & 0.85 & 0.73 & 0.81 & 0.90 & 0.94 \\
$L_4$  & 0.88 & 0.93 & ---  & 1.00 & ---  & 1.00 & ---  & 1.00 & 0.87 & 0.98 \\
$L_5$  & 0.91 & 0.99 & ---  & 1.00 & 0.86 & 0.97 & 0.82 & 0.93 & 0.93 & 1.00 \\
$L_6$  & 1.00 & 1.00 & 0.94 & 1.00 & 0.97 & 0.94 & 0.97 & 0.98 & 1.00 & 1.00 \\
$L_7$  & ---  & 1.00 & 0.77 & 0.96 & 0.77 & 0.98 & 1.00 & 1.00 & 1.00 & 1.00 \\
$L_8$  & 0.94 & 0.99 & 0.84 & 0.90 & 0.82 & 0.85 & 0.89 & 0.95 & 0.89 & 0.92 \\
$L_9$  & 0.81 & 0.87 & ---  & 1.00 & 0.97 & 0.97 & 0.90 & 0.96 & 0.82 & 0.92 \\
$L_{10}$ & 0.90 & 0.96 & 1.00 & 1.00 & 1.00 & 1.00 & 1.00 & 1.00 & ---  & 1.00 \\
$L_{11}$ & 0.83 & 0.88 & 0.85 & 0.90 & 0.80 & 0.92 & 0.71 & 0.87 & 0.85 & 0.95 \\
$L_{12}$ & 0.75 & 0.99 & 1.00 & 1.00 & 0.95 & 0.99 & 1.00 & 1.00 & 0.94 & 0.99 \\
$L_{13}$ & ---  & 1.00 & 0.97 & 1.00 & 0.86 & 0.99 & 0.98 & 0.99 & 0.89 & 0.99 \\ \bottomrule
\end{tabular}
\end{table}



\begin{table}[htbp!]
\footnotesize
\setlength{\tabcolsep}{3.5pt}
\caption{Linguistic feature analysis for each platform. Range: max-min value for each feature. Abbreviations: App = App Reviews, Code = Code Reviews, GH = GitHub, Jira = Jira, SO = Stack Overflow.}
\label{tab:statanalysis}
\begin{tabular}{@{}lrrrrrr@{}}
\toprule
\textbf{Linguistic Feature} & \textbf{App} & \textbf{Code} & \textbf{GH} & \textbf{Jira} & \textbf{SO} & \textbf{Range} \\ \midrule
Direct emotion                  & 60.6\% & 7.4\%  & 18.9\% & 12.8\% & 11.3\% & 53.2\% \\
Emphasized positivity           & 31.3\% & 0.9\%  & 4.3\%  & 5.6\%  & 7.7\%  & 30.5\% \\
Technical focus                 & 0.0\%  & 49.4\% & 45.6\% & 67.6\% & 70.1\% & 70.1\% \\
Balanced critique               & 31.9\% & 0.0\%  & 0.0\%  & 0.0\%  & 4.8\%  & 31.9\% \\
Progress sharing                & 2.1\%  & 0.0\%  & 8.4\%  & 15.0\% & 1.3\%  & 15.0\% \\
Gratitude                       & 8.1\%  & 2.6\%  & 35.1\% & 57.8\% & 15.4\% & 55.2\% \\
Inquisitive                     & 0.0\%  & 6.8\%  & 3.8\%  & 0.3\%  & 0.3\%  & 6.8\%  \\
Help Seeking and Offering       & 3.1\%  & 28.7\% & 41.5\% & 17.7\% & 49.9\% & 46.8\% \\
Compliments                     & 63.6\% & 0.0\%  & 16.0\% & 20.0\% & 18.7\% & 63.6\% \\
Bug Fix Requests                & 17.6\% & 0.3\%  & 0.5\%  & 0.5\%  & 0.0\%  & 17.6\% \\
Constructive Criticism          & 66.7\% & 37.7\% & 14.6\% & 20.0\% & 12.7\% & 54.0\% \\
Username Mentioning             & 0.5\%  & 0.6\%  & 11.9\% & 1.4\%  & 4.6\%  & 11.3\% \\
Name Mentioning                 & 0.0\%  & 3.9\%  & 1.9\%  & 20.7\% & 3.2\%  & 20.7\% \\ \bottomrule
\end{tabular}
\end{table}

Table~\ref{tab:statanalysis} shows considerable variation in linguistic features across platforms. For instance, features like \textit{Direct emotion}, \textit{Technical focus}, and \textit{Compliments} show wide ranges (over 50 percentage points), while other features (e.g., \textit{Inquisitive}, \textit{Username Mentioning}) exhibit less variation. Thus, some features are highly platform-specific, while others are broadly shared.

\vspace{3pt}
\setlength{\shadowsize}{2pt}
\noindent
\shadowbox{
\begin{minipage}{0.94\columnwidth}
	\textbf{Observation:} While platforms differ in several linguistic features, similarities remain for certain aspects.
\end{minipage}
}

\subsection{Statistical Analysis}
In addition to linguistic features, we examined eight statistical attributes per platform (see Table~\ref{tab:statisticanalysis}). 

\begin{table}[htbp!]
\footnotesize
\caption{Statistical analysis for each platform (averages per document).}
\label{tab:statisticanalysis}
\begin{tabular}{@{}lrrrrr@{}}
\toprule
\textbf{Statistic} & \textbf{App} & \textbf{Code} & \textbf{GH} & \textbf{Jira} & \textbf{SO} \\ \midrule
Characters               & 176.36 & 165.74 & 165.55 & 104.21 & 144.39 \\
Characters per word      & 4.14   & 4.71   & 4.53   & 4.61   & 4.57   \\
Words                    & 33.15  & 28.86  & 27.78  & 17.43  & 25.46  \\
Capitalized words        & 0.00   & 0.37   & 0.44   & 0.22   & 0.50   \\
Spelling mistakes        & 0.76   & 1.98   & 2.97   & 1.06   & 1.38   \\
Emoticons per document   & 0.02   & 0.07   & 0.40   & 0.14   & 0.18   \\
Question marks           & 0.29   & 0.32   & 0.24   & 0.14   & 0.22   \\
Exclamation marks        & 0.32   & 0.03   & 0.48   & 0.21   & 0.27   \\ \bottomrule
\end{tabular}
\end{table}

Distinct differences appear for statistics such as average number of characters (Jira shows shorter entries) and spelling mistakes (highest in GH). Other measures, such as characters per word, are nearly identical across all platforms. 

\vspace{3pt}
\setlength{\shadowsize}{2pt}
\noindent
\shadowbox{
\begin{minipage}{0.94\columnwidth}
	\textbf{Observation:} Platforms exhibit clear deviations in some statistics, but also share similarities, echoing findings from the linguistic analysis.
\end{minipage}
}

\subsection{Performance Analysis}

Table~\ref{tab:performances} reports the performance of the 14 sentiment analysis tools across 10 datasets, using micro/macro $F_1$ and their average (see Section~\ref{chap:metrics}). The highest value per dataset and metric is in bold.

\begin{table*}[htbp!]
\footnotesize
\setlength{\tabcolsep}{4pt}
\caption{Performance of each tool for each dataset. The best value for each metric and dataset is highlighted in bold.}
\label{tab:performances}
\begin{tabular*}{\textwidth}{@{\extracolsep{\fill}} llrrrrrrrrrrrrr}
\toprule
\rotatebox{0}{\textbf{Dataset}}                & \rotatebox{0}{\textbf{Metrics}}             & \rotatebox{90}{\textbf{ALBERT}}        & \rotatebox{90}{\textbf{BERT}} & \rotatebox{90}{\textbf{DistilBERT}}    & \rotatebox{90}{\textbf{ELECTRA}} & \rotatebox{90}{\textbf{RoBERTa}}       & \rotatebox{90}{\textbf{SetFit}}        & \rotatebox{90}{\textbf{XLNet}}  & \rotatebox{90}{\textbf{Senti4SD}} & \rotatebox{90}{\textbf{SEnti-Analyzer}} & \rotatebox{90}{\textbf{SentiCR}} & \rotatebox{90}{\textbf{SentiStrength}} & \rotatebox{90}{\textbf{SentiStrength-SE}} & \rotatebox{90}{\textbf{SentiSW}} \\ \midrule
\multirow{3}{*}{App}    & Micro   averaged F1 & 0.91          & 0.93 & 0.93          & 0.75    & 0.91          & \textbf{0.94} & 0.93  & 0.87     & 0.59                                                      & 0.74    & 0.62          & 0.59             & 0.80    \\
                        & Macro   averaged F1 & 0.61          & 0.62 & 0.63          & 0.50    & 0.63          & \textbf{0.64} & 0.63  & 0.57     & 0.53                                                      & 0.52    & 0.49          & 0.50             & 0.56    \\
                        & Overall   score     & 0.76          & 0.78 & 0.78          & 0.63    & 0.77          & \textbf{0.79} & 0.78  & 0.72     & 0.56                                                      & 0.63    & 0.55          & 0.54             & 0.68    \\ \midrule
\multirow{3}{*}{Code}   & Micro   averaged F1 & 0.87          & 0.87 & 0.89          & 0.86    & 0.86          & \textbf{0.89} & 0.82  & 0.81     & 0.77                                                      & 0.79    & 0.70          & 0.76             & 0.78    \\
                        & Macro   averaged F1 & 0.55          & 0.80 & 0.83          & 0.80    & 0.82          & \textbf{0.86} & 0.78  & 0.74     & 0.55                                                      & 0.72    & 0.58          & 0.58             & 0.69    \\
                        & Overall   score     & 0.71          & 0.83 & 0.86          & 0.83    & 0.84          & \textbf{0.87} & 0.80  & 0.77     & 0.66                                                      & 0.75    & 0.64          & 0.67             & 0.74    \\ \midrule
\multirow{3}{*}{GH 1}   & Micro   averaged F1 & 0.92          & 0.92 & 0.92          & 0.92    & 0.93          & \textbf{0.94} & 0.92  & 0.89     & 0.72                                                      & 0.82    & 0.61          & 0.70             & 0.78    \\
                        & Macro   averaged F1 & 0.92          & 0.92 & 0.92          & 0.92    & 0.93          & \textbf{0.94} & 0.92  & 0.90     & 0.69                                                      & 0.81    & 0.62          & 0.68             & 0.79    \\
                        & Overall   score     & 0.92          & 0.92 & 0.92          & 0.92    & 0.93          & \textbf{0.94} & 0.92  & 0.89     & 0.71                                                      & 0.81    & 0.62          & 0.69             & 0.79    \\ \midrule
\multirow{3}{*}{GH 2}   & Micro   averaged F1 & 0.84          & 0.81 & \textbf{0.84} & 0.80    & 0.82          & 0.83          & 0.76  & 0.75     & 0.72                                                      & 0.74    & 0.52          & 0.67             & 0.74    \\
                        & Macro   averaged F1 & 0.73          & 0.72 & \textbf{0.77} & 0.67    & 0.72          & 0.75          & 0.70  & 0.61     & 0.57                                                      & 0.65    & 0.52          & 0.60             & 0.62    \\
                        & Overall   score     & 0.78          & 0.76 & \textbf{0.81} & 0.74    & 0.77          & 0.79          & 0.73  & 0.68     & 0.64                                                      & 0.69    & 0.52          & 0.65             & 0.68    \\ \midrule
\multirow{3}{*}{GH 3}   & Micro   averaged F1 & 0.69          & 0.71 & \textbf{0.75} & 0.71    & 0.72          & 0.74          & 0.67  & 0.71     & 0.72                                                      & 0.68    & 0.52          & 0.69             & 0.67    \\
                        & Macro   averaged F1 & 0.44          & 0.28 & 0.48          & 0.28    & 0.52          & \textbf{0.58} & 0.45  & 0.41     & 0.54                                                      & 0.55    & 0.50          & 0.54             & 0.47    \\
                        & Overall   score     & 0.57          & 0.49 & 0.61          & 0.49    & 0.62          & \textbf{0.66} & 0.56  & 0.56     & 0.63                                                      & 0.61    & 0.51          & 0.61             & 0.57    \\ \midrule
\multirow{3}{*}{Jira 1} & Micro   averaged F1 & 0.95          & 0.94 & 0.96          & 0.96    & 0.96          & \textbf{0.97} & 0.94  & 0.88     & 0.79                                                      & 0.87    & 0.74          & 0.83             & 0.86    \\
                        & Macro   averaged F1 & 0.94          & 0.94 & 0.96          & 0.96    & 0.96          & \textbf{0.96} & 0.93  & 0.88     & 0.78                                                      & 0.85    & 0.74          & 0.83             & 0.86    \\
                        & Overall   score     & 0.94          & 0.94 & 0.96          & 0.96    & 0.96          & \textbf{0.97} & 0.93  & 0.88     & 0.78                                                      & 0.86    & 0.74          & 0.83             & 0.86    \\ \midrule
\multirow{3}{*}{Jira 2} & Micro averaged F1   & \textbf{0.87} & 0.85 & 0.84          & 0.86    & 0.86          & 0.85          & 0.86  & 0.81     & 0.80                                                      & 0.85    & 0.65          & 0.80             & 0.84    \\
                        & Macro   averaged F1 & \textbf{0.82} & 0.80 & 0.79          & 0.81    & 0.81          & 0.80          & 0.82  & 0.77     & 0.74                                                      & 0.78    & 0.69          & 0.79             & 0.80    \\
                        & Overall   score     & \textbf{0.84} & 0.82 & 0.81          & 0.84    & 0.84          & 0.82          & 0.84  & 0.79     & 0.77                                                      & 0.81    & 0.67          & 0.79             & 0.82    \\ \midrule
\multirow{3}{*}{SO 1}   & Micro   averaged F1 & 0.89          & 0.86 & 0.89          & 0.85    & 0.89          & \textbf{0.91} & 0.90  & 0.86     & 0.80                                                      & 0.83    & 0.69          & 0.78             & 0.83    \\
                        & Macro   averaged F1 & 0.79          & 0.71 & 0.77          & 0.48    & 0.79          & \textbf{0.80} & 0.79  & 0.70     & 0.46                                                      & 0.66    & 0.54          & 0.42             & 0.64    \\
                        & Overall   score     & 0.84          & 0.79 & 0.83          & 0.67    & 0.84          & \textbf{0.85} & 0.84  & 0.78     & 0.62                                                      & 0.75    & 0.61          & 0.60             & 0.73    \\ \midrule
\multirow{3}{*}{SO 2}   & Micro   averaged F1 & 0.84          & 0.82 & 0.84          & 0.85    & \textbf{0.87} & 0.84          & 0.85  & 0.82     & 0.83                                                      & 0.80    & 0.76          & 0.77             & 0.79    \\
                        & Macro   averaged F1 & 0.83          & 0.81 & 0.83          & 0.84    & \textbf{0.86} & 0.83          & 0.85  & 0.81     & 0.83                                                      & 0.78    & 0.79          & 0.78             & 0.78    \\
                        & Overall   score     & 0.83          & 0.81 & 0.84          & 0.84    & \textbf{0.87} & 0.84          & 0.85  & 0.81     & 0.83                                                      & 0.79    & 0.78          & 0.78             & 0.68    \\ \midrule
\multirow{3}{*}{SO 3}   & Micro   averaged F1 & 0.90          & 0.88 & 0.89          & 0.86    & 0.90          & \textbf{0.91} & 0.85  & 0.83     & 0.74                                                      & 0.80    & 0.65          & 0.73             & 0.80    \\
                        & Macro   averaged F1 & 0.83          & 0.81 & 0.85          & 0.77    & 0.84          & \textbf{0.86} & 0.77   & 0.76     & 0.56                                                      & 0.64    & 0.56          & 0.52             & 0.56    \\
                        & Overall   score     & 0.87          & 0.85 & 0.87          & 0.81    & 0.87          & \textbf{0.89} & 0.81   & 0.80     & 0.60                                                      & 0.72    & 0.60          & 0.63             & 0.68    \\ \bottomrule
\end{tabular*}
\end{table*}

Across most datasets and metrics, SetFit achieves the highest overall score, with RoBERTa and other BERT-based transformer models also performing strongly. For Stack Overflow datasets, SetFit and RoBERTa achieve nearly identical results. Notably, the software engineering-specific tools (e.g., Senti4SD, SentiStrength-SE) do not outperform transformer models in any scenario.

\vspace{3pt}
\setlength{\shadowsize}{2pt}
\noindent
\shadowbox{
\begin{minipage}{0.94\columnwidth}
	\textbf{Observation:} Transformer-based models such as SetFit and RoBERTa show robust performance across all datasets. SE-specific tools do not outperform these models on any dataset.
\end{minipage}
}

\subsection{Questionnaire Mapping}

Based on the results from Table~\ref{tab:statanalysis} in Section~\ref{chap:characteristicanalysis}, we mapped each platform to one of the four intervals described in Section~\ref{chap:questionnaire}. Table~\ref{tab:mapping} presents this mapping for each linguistic feature.

\begin{table}[htbp!]
\footnotesize
\setlength{\tabcolsep}{4pt}
\caption{Mapping of the platforms to the four answer options for each linguistic feature}
\label{tab:mapping}
\begin{tabularx}{\columnwidth}{rlXXX}
\toprule
\textbf{$L_i$ } & \textbf{True} & \textbf{Likely}   & \textbf{Unlikely} & \textbf{Untrue} \\ \midrule
$L_1$                   & -             & App               & -                 & Code, GH, Jira, SO \\
$L_2$                   & -             & -                 & App               & Code, GH, Jira, SO \\
$L_3$                   & -             & Jira, SO           & Code, GH           & App             \\
$L_4$                   & -             & -                 & App               & Code, GH, Jira, SO \\
$L_5$                   & -             & -                 & -                 & All             \\
$L_6$                   & -             & Jira              & GH                & App,Code, SO     \\
$L_7$                   & -             & -                 & -                 & All             \\
$L_8$                   & -             & -                 & Code, GH, SO        & App, Jira        \\
$L_9$                   & -             & App               & -                 & Code, GH, Jira, SO \\
$L_{10}$                & -             & -                 & -                 & All             \\
$L_{11}$                & -             & App               & Code              & GH, Jira, SO      \\
$L_{12}$                & -             & -                 & -                 & All             \\
$L_{13}$                & -             & -                 & -                 & All             \\ \bottomrule
\end{tabularx}
\end{table}

Here, ``All'' means every platform falls into the answer option for that feature; ``-'' indicates no platform matches, resulting in an ambiguous assignment for that feature. As described in Section~\ref{chap:questionnaire}, a point is assigned to the ambiguous dataset if the user selects ``Not specified''. In the upper two intervals (``True'' and ``Likely''), ambiguous assignment is most common.

Table~\ref{tab:userinput} gives an example: User input is mapped to intervals, and for each platform it is checked whether it falls into the corresponding interval for that feature. Points are assigned accordingly.

\begin{table}[htbp!]
\footnotesize
\setlength{\tabcolsep}{4pt}
\caption{Example of user input for communication linguistic features. The platform \textit{GitHub} receives the highest score (9) and is thus suggested as the output.}
\label{tab:userinput}
\begin{tabularx}{\columnwidth}{@{}lXrrrrrrr@{}}
\toprule
\textbf{$L_i$} & \textbf{User Input}  & \textbf{Interval} & \textbf{App} & \textbf{Code} & \textbf{GH} & \textbf{Jira} & \textbf{SO} & \textbf{Ambig} \\ \midrule
$L_{1}$                & Untrue        & 0-25\%   & 0   & 1    & 1          & 1    & 1  & 0         \\
$L_{2}$               & Untrue        & 0-25\%   & 0   & 1    & 1          & 1    & 1  & 0         \\
$L_{3}$              & Unlikely      & 25-50\%  & 0   & 1    & 1          & 0    & 0  & 0         \\
$L_{4}$               & Untrue        & 0-25\%   & 0   & 1    & 1          & 1    & 1  & 0         \\
$L_{5}$               & Unlikely      & 25-50\%  & 0   & 0    & 0          & 0    & 0  & 1         \\
$L_{6}$               & Unlikely      & 25-50\%  & 0   & 0    & 1          & 0    & 0  & 0         \\
$L_{7}$               & Untrue        & 0-25\%   & 1   & 1    & 1          & 1    & 1  & 0         \\
$L_{8}$               & Unlikely      & 25-50\%  & 0   & 1    & 1          & 0    & 1  & 0         \\
$L_{9}$              & Untrue        & 0-25\%   & 0   & 1    & 1          & 1    & 1  & 0         \\
$L_{10}$              & N/A & -        & 0   & 0    & 0          & 0    & 0  & 1         \\
$L_{11}$              & Untrue        & 0-25\%   & 0   & 0    & 1          & 1    & 1  & 0         \\
$L_{12}$              & N/A  & -        & 0   & 0    & 0          & 0    & 0  & 1         \\
$L_{13}$              & Unlikely      & 25-50\%  & 0   & 0    & 0          & 0    & 0  & 1         \\ \midrule
\textbf{Sum}             & -             & -        & 1   & 7    & \textbf{9} & 6    & 7  & 4         \\ \bottomrule
\end{tabularx}
\end{table}

The platform with the most points is suggested as the most similar to the user’s dataset; in this example, \textit{GitHub}. If no clear match exists, or “Not specified” dominates, the ambiguous dataset is returned, and robust cross-platform tools such as SetFit or SentiStrength-SE are suggested.

\vspace{3pt}
\setlength{\shadowsize}{2pt}
\noindent
\shadowbox{
\begin{minipage}{0.94\columnwidth}
	\textbf{Observation:} The mapping enables identification of the most similar platform and the corresponding recommended tool for a user's dataset.
\end{minipage}
}

\section{Discussion}
\label{chap:discussion}

\subsection{Answering the Research Questions}
Based on the results from Section~\ref{chap:results}, we summarize our research questions as follows:

\noindent\textbf{RQ1:} There are pronounced differences in linguistic features and statistics across platforms, though some features (e.g., average word length) are similar. Every linguistic feature shows at least a two-digit range between platforms, indicating that platform-specific communication styles are distinct and measurable.

\noindent\textbf{RQ2:} SetFit~\cite{tunstall2022setfit} achieved the best overall performance for most datasets, with RoBERTa and ELECTRA also leading for certain platforms. However, ``best-performing'' is highly context-dependent; tool effectiveness varies with dataset characteristics such as communication style, class balance, and domain-specific language. Our results reflect relative performance on the selected datasets and may not generalize to other SE contexts or future datasets.

\noindent\textbf{RQ3:} By mapping dataset characteristics to user input via the questionnaire, users can identify which platform is most similar to their data, enabling selection of a tool that has shown strong performance for comparable datasets.

\subsection{Interpretation and Implications}

The proposed approach distinguishes platforms based on both linguistic features and simple statistics. Features such as ``Gratitude'' (occurring in 57.8\% of Jira data versus 2.6\% in Code Reviews) demonstrate that both quantitative and qualitative dimensions are necessary to meaningfully differentiate SE datasets. Nevertheless, some features (e.g., average characters per word, number of question marks) exhibit little variance and may be omitted or combined in future iterations to streamline analysis.

Performance results reinforce that transformer-based models (SetFit, RoBERTa, ELECTRA) consistently outperform SE-specific or lexicon-based tools, in line with recent advances in trustworthy AI and requirements engineering. However, performance varies not just by platform but even within platforms, suggesting that tool selection should account for additional factors such as data source (e.g., pull requests vs. issue comments), dataset size, and labeling practices. While we report $F_1$-score as the main metric, users with different priorities (e.g., high recall for negative cases) may complement this analysis with other metrics to match their specific trustworthiness or risk management requirements.

The prevalence of ``Ambiguous'' mappings in the questionnaire indicates that further refinement is needed, either through improved feature selection, normalization, or more granular groupings that move beyond platform labels toward characteristic profiles. Future work may use clustering or embedding-based similarity measures to automate dataset matching and support evolving requirements for trustworthy and explainable AI systems.

Cross-platform performance gaps remain a challenge: even state-of-the-art tools perform worse on datasets that differ from their training data. A detailed correlation analysis between dataset characteristics and tool accuracy, as well as robust statistical testing of performance differences, could further substantiate these findings and provide actionable recommendations for requirements engineering in AI-based systems.

In practical terms, for direct, technical discussions (e.g., GitHub issues), lexicon-based tools may suffice due to their stability. For complex or emotionally nuanced contexts (e.g., Jira, team chats), fine-tuned machine learning models are preferable. Ultimately, selecting a tool should align with the project’s requirements for trustworthiness, explainability, and adaptability. Additionally, ensemble or voting-based approaches, or even newer large language models (LLMs), could be explored as alternatives, though these were not directly evaluated in this study.

\subsection{Threats to Validity}
\label{sec:threats}

We categorize the threats to our analysis following Wohlin et al.~\cite{wohlin2012experimentation}.

\textit{Construct Validity.}  
Our analysis is limited to specific projects from each platform, and alternative categorizations (e.g., by project type or granularity) could yield different results. We adopted platform definitions consistent with previous studies~\cite{obaidi22cross,9240704,noviellicross20}. When mapping user input for statistics in the questionnaire, we used a distance measure, which does not allow for ambiguous assignment in absolute terms; setting empirical thresholds would be subjective, as no standard exists. Hyperparameters for sentiment tools were not tuned per dataset, nor were advanced analyses (e.g., k-fold cross-validation) performed, as our aim was to demonstrate the overall approach rather than optimize for marginal gains. The set of statistics chosen may not cover all possible correlations to platform characteristics, but was guided by prior work~\cite{10.1007/978-3-030-64266-2_8}, and our method allows for statistics to be adapted flexibly.

The SEnti-Analyzer tool uses a majority-voting ensemble, which may behave differently from single tools; however, prior work shows only minor differences~\cite{obaidi22cross,uddinVoting2021}.  
A further threat is the inherent subjectivity of sentiment interpretation in SE: even gold-standard labels reflect annotator judgments that may not generalize~\cite{herrmann2025different-perceptions,herrmannSentiSurvey22}. This could impact our performance comparisons and recommendations.

\textit{Internal Validity.}  
For platform analysis, datasets were merged and sampled; differences between datasets from the same platform or varying crawling methods could introduce bias, though this risk is mitigated by source consistency. Performance was analyzed per dataset and platform. Use of third-party tools (e.g., pyspellchecker) inherits their limitations.

\textit{Conclusion Validity.}  

Original datasets may have been filtered in ways that affect statistical analysis; we selected statistics likely unaffected by such filters. Coding of linguistic features remains partly subjective, despite calibration workshops among researchers. The identification of a “best-performing” tool is relative to each dataset, and results may not generalize beyond the contexts evaluated. Future work may include additional validation metrics or risk-weighted performance measures to address a broader range of application scenarios.

\textit{External Validity.}  

Results are specific to the sampled data and platforms analyzed and may not generalize to all SE communication or future datasets. To reduce this threat, we included all available sentiment datasets for each platform to capture diversity within current data. The generalizability to new data types or to datasets with different communication profiles should be further evaluated in future studies.

\subsection{Future Work}

Future research could extend the approach to broader social and behavioral aspects of software engineering, for example by clustering questionnaire responses to gain deeper insights into team and communication dynamics. The subjectivity and inconsistency in sentiment labeling, as highlighted in prior studies~\cite{10.1145/3194932.3194935,Imtiaz.2018,M.Ortu.2015,Murgia.2014,CABRERADIEGO2020105633,8643972,herrmannSentiSurvey22,obaidi2022sentisurveyzenodo}, underline the value of standardized annotation processes, such as those proposed by Sun et al.~\cite{sun2021}, potentially leveraging established emotion models~\cite{shaver1987emotion,parrott2001emotions}. Applying our methodology to datasets labeled under such standards could improve consistency and enable more robust cross-platform analyses.

Collaborations with industry partners could enable the inclusion of proprietary datasets, facilitating comparison between industrial and public communication channels, and covering platforms like Microsoft Teams, Slack, or Discord. The approach could also be adapted to domains beyond sentiment analysis, such as software-specific datasets.

Further enhancements may include evaluating usability and software quality aspects of sentiment analysis tools, performing correlation analyses between dataset characteristics and tool performance, or refining mapping scales (e.g., using three-point rather than four-point scales). Automating the identification of linguistic features with machine learning or clustering techniques (e.g., k-means) could make the mapping process more objective and efficient. Prioritizing features based on their predictive power for tool performance could also improve recommendation accuracy.

Importantly, the questionnaire should be validated in practice to assess its utility and accuracy in real-world scenarios. Additional work could explore a broader range of NLP statistics, focus analyses on individual datasets, and examine the influence of contextual factors such as project domain, team composition, or communication culture. These aspects may be investigated through real-world case studies or controlled user studies to further refine and validate tool recommendations.  Finally, future work should include evaluations of large language models (LLMs), ensemble methods, and additional evaluation metrics (such as recall, precision, or risk-based measures) to ensure relevance for new requirements engineering and trustworthy AI scenarios.

\section{Conclusion}
\label{chap:conclusion}

Sentiment analysis is an increasingly relevant technique in software engineering, especially for understanding collaborative communication and supporting trustworthy AI-based systems. However, our results show that sentiment analysis tool performance varies considerably depending on the linguistic and statistical characteristics of the dataset—particularly in cross-platform scenarios. Tools trained on one type of developer communication, such as GitHub, often perform suboptimally when applied to different settings like Jira due to diverse communication styles and content structures.

To address this, we presented a systematic approach that analyzes dataset characteristics and matches them with the most suitable sentiment analysis tools. By evaluating 14 tools across 10 datasets, we identified substantial differences in linguistic features and statistics between platforms. Based on these findings, we developed a questionnaire-based mapping system that allows users to describe their dataset and receive evidence-based tool recommendations—without requiring labeled data.

Our results confirm that transformer-based models such as SetFit, RoBERTa, and ELECTRA consistently achieve top performance, but also highlight that no tool is universally optimal. Tool effectiveness remains highly context-dependent, underscoring the need for careful tool selection in practice. As software engineering moves toward greater use of AI and automated analytics, aligning sentiment analysis with platform-specific characteristics is crucial for reliable and trustworthy insights.

Future research should further investigate the influence of contextual and domain-specific variables on tool performance, and evaluate the practical impact of recommendation systems like ours in real-world software projects and requirements engineering for AI. 

\section*{Data Availability Statements}
The raw dataset containing the labeling result is available on \href{https://doi.org/10.6084/m9.figshare.29250935.v1}{figshare} \cite{obaidi2025datasetretrai}.


\bibliographystyle{IEEEtran}
\bibliography{references.bib}

\end{document}